\documentclass{aa} 

\usepackage{graphicx}
\usepackage[varg]{txfonts}
\usepackage{natbib}
\usepackage{lscape}
\usepackage[font={small,rm}]{caption}
\usepackage{array}
\usepackage{bm}
\usepackage{color}
\usepackage[T1]{fontenc}
\usepackage{multirow}

\begin{document} 
   \title{The mid-IR water and silicate relation in protoplanetary disks}
\authorrunning{Antonellini et al.}
   \subtitle{}

\author{Antonellini$^{1}$, S. \and Bremer, J.$^{1}$ \and Kamp, I.$^{1}$ \and Riviere-Marichalar, P.$^{2}$ \and Lahuis, F.$^{1,3}$ \and Thi, W.-F.$^{4}$
\and Woitke, P.$^{5}$ \and Meijerink, R.$^{6}$ \and Aresu, G.$^{1,7}$ \and Spaans, M.$^{1}$}

	   \institute{Kapteyn Astronomical Institute,
		      Postbus 800, 9700 AV Groningen, The Netherlands\\
		      \email{antonellini@astro.rug.nl}\and
		      Centro de Astrobiolog\'ia (INTA-CSIC) - Depto. Astrof\'isica, POB 78, ESAC Campus, 28691 Villanueva de la Ca\~nada, Spain\and
		      SRON Netherlands Institute for Space Research, P.O. Box 800, 9700 AV Groningen, The Netherlands\and
		      Max-Planck-Institut f$\rm\ddot{u}$r extraterrestrische Physisk, Giessenbachstrasse 1, 85748 Garching, Germany\and
		      St. Andrews University, School of Physics and Astronomy\and
		      Leiden Observatory, Leiden University, PO Box, 2300 RA Leiden, The Netherlands\and
		      INAF, Osservatorio Astronomico di Cagliari, via della Scienza 5, 09047 Selargius, Italy}

	   \date{}

	 
	  \abstract
{Mid-IR water lines from protoplanetary disks around T Tauri stars have a detection rate of 50\%. Models have identified multiple physical properties of disks such as dust-to-gas mass ratio, dust size power law 
distribution, disk gas mass, disk inner radius, and disk scale height as potential explanation for the current detection rate.}
{In this study, we aim to break degeneracies through constraints obtained from observations. We search for a connection between mid-IR water line fluxes and the strength of the 10~$\mu$m silicate feature.}
{We analyse observed water line fluxes from three blends at 15.17, 17.22, 29.85~$\mu$m published earlier and compute the 10~$\mu$m silicate feature strength from Spitzer spectra to search for possible trends. We use a series of 
published ProDiMo thermo-chemical models, exploring disk dust and gas properties, and the effects of different central stars. In addition, we produced two standard models with different dust opacity functions,
and one with a parametric prescription for the dust settling.} 
{Our model series that vary properties of the grain size distribution suggest that mid-IR water emission anticorrelates with the strength of the 10~$\mu$m silicate feature. The models also show that the 
increasing stellar bolometric luminosity enhance simultaneously the strength of this dust feature and the water lines fluxes.
No correlation is found between the observed mid-IR water lines and the 10~$\mu$m silicate strength. 2/3 of targets in our sample show crystalline dust features, and the disks are mainly flaring. Our sample shows the same 
difference in the peak strength between amorphous and crystalline silicates that was noted in earlier studies, but our models do not support this intrinsic difference in silicate peak strength. Individual properties of our model 
series are not able to reproduce the most extreme observations, suggesting that more complex (e.g. vertically changing) dust properties are required to reproduce the strongest 10~$\mu$m silicate features. A parametrized settling 
prescription is able to boost the peak strength by a factor 2 for the standard model. Water line fluxes are unrelated to the composition of the dust.
The pronounced regular trends seen in the model results are washed out in the data due to the larger diversity in stellar and disk properties compared to our model series.} 
{The independent nature of the water line emission and the 10~$\mu$m silicate strength found in observations, and the modeling results, leave as a possible explanation that the disks with with weaker mid-IR 
water line fluxes are depleted in gas or enhanced in dust in the inner 10~au. In the case of gas depleted disks, settling produces very strong 10~$\mu$m silicate features, with strong peak strength. Observations of larger 
unbiased samples with JWST/MIRI and ALMA are essential to verify this hypothesis.}


{}

{}


	   \maketitle
	%

\section{Introduction}\vspace{3mm}

Water has been observed in protoplanetary disks from near-IR \citep{salyk}, through mid-IR \citep{carr,pontoppidan1,pontoppidan2}, to far-IR \citep{hogerheijde,podio,riviere-marichalar}. The main water reservoir, with an 
abundance of 10$^{-4}$ with respect to the total hydrogen number density, is located in the inner few au; this reservoir is responsible for water emission below 30~$\mu$m. A survey of mid-IR water lines produced a complex 
picture, with a detection rate of about 50\% for T~Tauri and 5\% for Herbig stars \citep{pontoppidan1}. Our previous work \citep{antonellini}, found that the continuum opacity due to differences in dust properties is one of the 
main reasons for changing the strength of the mid-IR water emission. 
In a subsequent paper \citep{antonellini1}, we explain the non-detections of mid-IR water lines around early type Pre Main-Sequence stars (PMSs), with the stronger mid-IR continuum and instrumental noise.

Dust is assumed to constitute only 1\% of the disk mass, but it is a bolometric source of opacity, able to suppress line fluxes. Dust also plays a role in the thermal balance of the disk, due to the photoelectric effect. In 
environments such as protoplanetary disks, dust grains grow \citep[as inferred from mm flux and spectral indexes,][]{beckwith,mannings,andrews2,andrews3} and settle \citep[][]{d'alessio1,chiang1}. Silicate grains produce two 
solid state features around 10 and 20~$\mu$m. The features originate mainly from $\mu$m-size grains where the dust temperature is around 200-600~K \citep{natta2}. The 10~$\mu$m feature hosts information about the thermal and 
physical history of dust grains, and also the composition \citep{kessler-silacci}. According to \citet{vanboekel}, the more the dust is processed (crystallized), the wider the feature and the lower the peak strength. The 
strength of the 10~$\mu$m peak is also inversely correlated with the dust grain size. \citet{bouwman2} found a correlation between the shape of the 10~$\mu$m amorphous feature and the SED shape, that is consistent with both dust 
growth and subsequent settling. Turbulent mixing however seems efficient in maintaining $\mu$m-sized grains in the atmospheres of T~Tauri disks \citep{olofsson}.
Mass transport can also explain the presence of a relevant fraction of cold crystallized dust in disks around T~Tauri stars \citep{olofsson1}. The correlations found between crystalline indexes and continuum spectral indexes 
considered in \citet{watson2}, are consistent with the fact that settled disks have a larger mass of crystalline silicates.
In our previous work \citep{antonellini}, we found the continuum opacity to be a main actor in IR water spectroscopy. The dust opacity depends on the size of the dust (both in terms of maximum size and size 
distribution) and its composition. Typical spectral features of silicates in mid-IR spectra are at $\sim$10$~\mu$m and $\sim$18$~\mu$m; the exact wavelength depends on the composition \citep{jager}. Also the shape and intensity 
of these features are related to the dust composition and size properties \citep[e.g.][]{jaeger,chiang1,bouwman1,min1}. This suggests the potential existence of a relation between the strength/shape of the strongest dust 
silicate feature (the 10$~\mu$m) and the mid-IR water lines. The presence of this correlation could open the possibility of a new diagnostic tool for the inner disk properties.
\nolinebreak

Both water lines and the silicate features arise from the inner disk. Hence, in this paper we investigate a potential relation between the mid-IR water lines and the 10~$\mu$m silicate feature strength. We investigate this 
in the context of the thermo-chemical modeling results by \citet{antonellini,antonellini1}. And we compared these findings to a sample of disks observed with Spitzer.

\section{Spitzer spectra}\vspace{3mm}
\label{s1}

We analyze in this work Spitzer-IRS high ($R~=~600$) and low resolution ($R~\lesssim~120$) spectra of a sample of T~Tauri stars. We consider T~Tauri stars for which measurements of line fluxes for the blends at 15.17~$\mu$m, 
17.22~$\mu$m and 29.85~$\mu$m, are available from \citet{pontoppidan1}. We then took spectra from CASSIS database\footnote{The Cornell Atlas of Spitzer/IRS Sources (CASSIS) is a product of the Infrared Science Center at Cornell University, supported by NASA and JPL.} and reduced BCD data from the 
Spitzer archive\footnote{http://irsa.ipac.caltech.edu/data/SPITZER/docs/irs/features/}. We reduce the second set of high resolution data following a pipeline developed by the c2d team \citep{evans}, and applying the PSF 
extraction procedure \citep{lahuis2007b,lahuis2007PhD}. The data are then post-processed, applying defringing \citep{lahuis2003} to remove residuals due to low-level fringes. Spectra from CASSIS are all low resolution, and 
already reduced following a pipeline described in detail in \citet{lebouteiller}. Both the pipelines produce spectra of comparable quality beside the different resolution.


The strength of the 10~$\mu$m Silicate feature $S_\mathrm{\lambda,peak}$ is computed as

\begin{equation}\label{(1)}
S_\mathrm{\lambda,peak} = \frac{F_\mathrm{\lambda,peak}-F_\mathrm{\lambda,cont}}{F_\mathrm{\lambda,cont}},
\end{equation}

\noindent
where $F_\mathrm{\lambda}$ is the flux at 10~$\mu$m. $F_\mathrm{\lambda,cont}$ is the local linearly fitted continuum. The error computation is taken from the covariance matrix of the fitting function\footnote{$\mathrm{http://docs.scipy.org/doc/scipy-0.14.0/reference/generated/}$ $\mathrm{scipy.optimize.curve\_fit.html}$}.

\begin{figure}[htpl!]
\centering 
\begin{minipage}[l]{0.5\textwidth}
\includegraphics[width=\textwidth]{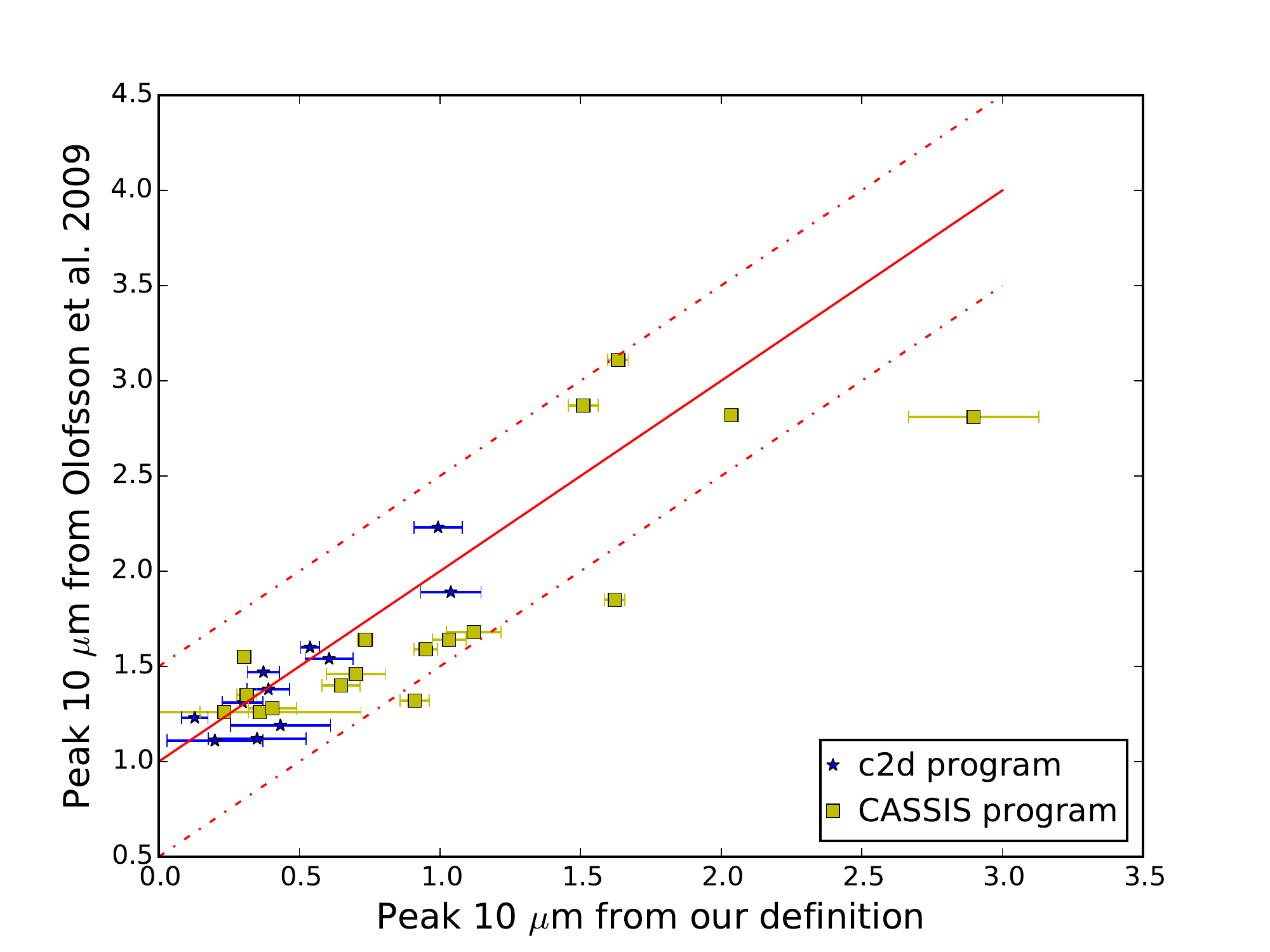}
\caption{Comparison of the peak strength definition adopted in our work and the one obtained by \citet{olofsson}. Blue stars are data extracted from c2d program Spitzer spectra \citep{evans}, while yellow squares
are data extracted from CASSIS \citep{lebouteiller} Spitzer spectra. The red solid line is the reference. Dot-dashed lines indicate 50\% of deviation from the 1 to 1 relation. 
\citet{olofsson} data are based on c2d spectra, and do not contain errors.} 
\label{Comp}
\end{minipage}
\end{figure}

\noindent
The definition of silicate peak strength adopted in this work, differs from the one used in \citet{olofsson}, which is based on \citet{kessler-silacci1}. There, the peak strength is defined as

\begin{equation}\label{(2)}
F_\mathrm{peak}\!=\!1\!+\!\frac{F_{\nu}-F_\mathrm{\nu,continuum}}{\langle F_\mathrm{continuum}\rangle},
\end{equation}

\noindent
where $F_{\nu}$ is the spectral flux at the frequency of the peak, $F_\mathrm{\nu,continuum}$ is the fitted value for the continuum at the peak frequency, and $\langle F_\mathrm{continuum}\rangle$ is the average continuum in the
fitted wavelength range $\sim$7-14~$\mu$m, the exact range of the fit is a function of the data range and SED slope. Beside the +1 shift in the value of the peak (that will systematically produce larger peak 
values with respect to our definition), the definition of the average continuum can produce a slight difference in the ratio, in particular in cases where our fitted range differs from the 
one adjusted for each specific case. 

Fig.~\ref{Comp} shows that our definition produces systematically weaker peak fluxes, but the deviations scatter from the red line is around one. In some cases, our selected spectra are from a different program, and some of the 
targets are known to be variable in the mid-IR. \citet{olofsson} fluxes are from c2d spectra, and their values agree better with our peak fluxes in the case where we used c2d spectra as well. 
Agreement between the two peak strength definitions is on average better than 13\%, meaning that any conclusion produced will be valid independently of the definition used for the silicate feature strength.

Individual targets, have different spectral types and luminosities; in our sample, the range of luminosity spans from 0.26 to 2.6 $L_{\odot}$. To make a direct comparison between the 
observations and our model series (that is built around a typical K7 star with $L_{*}\!=\!0.7~$L$_{\odot}$), we divided the mid-IR water line fluxes by the local continuum flux.

\section{Modeling water \& silicates}\vspace{3mm}
\label{m2}

We perform a parameter space exploration using a sub-series of the models presented in \citet{antonellini,antonellini1}, considering parameters significantly affecting mid-IR water line fluxes and dust peak 
strength at 10~$\mu$m (Table~\ref{models}). These are 2D radiation thermo-chemical models, with continuous disk structure, and non-LTE line radiative transfer. The series are built by changing a single 
parameter each time. These models include a prescription of the dust settling based on \citet{dubrulle}. All our model series disk dust composition is based on a mixture of 80\% amorphous silicates and 20\% 
carbonaceous dust species with average $\rho\!=\!3.5~$g/cm$^3$ \citep[][]{draine2}, and the dust size distribution is based on a power law extending from a minimum (fixed for all the models $a_\mathrm{min}\!=\!0.05~\mu$m) 
to a maximum size (listed in Table~\ref{models}). From these models, we produced high resolution Spitzer spectra, and we extract the water blend fluxes at 15.17, 17.22 and 29.85~$\mu$m adopting the same procedure as described in 
\citet{pontoppidan1}. 
Finally we also computed the peak strength of the 10~$\mu$m silicate feature using Eq.~(\ref{(1)}).

\begin{table}
\centering
\begin{minipage}[l]{0.5\textwidth}
\caption{Disk parameters series}
\begin{tabular}{p{0.8cm} p{1.7cm} p{1.35cm} p{3.6cm}}
\hline\hline
\small{Model series} & \small{Parameter} & \small{Symbol} & \small{Series values/model}\\ \hline
1 & \small{Maximum dust size} & \small{$a_\mathrm{max}$ [$\mu$m]} & \small{250, 400, 500, 700, $\bm{1000}$, 2000, 5000, 10$^{4}$, 10$^{5}$}\\
2 & \small{Dust size power law index} & \small{$a_\mathrm{pow}$} & \small{2.0, 2.5, 3.0, $\bm{3.5}$, 4.0, 4.5}\\
3 & \small{Disk gas mass} & \small{$M_\mathrm{gas}$ [M$_\mathrm{\odot}$]} & \small{10$^{-5}$, 10$^{-4}$, 0.001, $\bm{0.01}$, 0.05, 0.1}\\
4 & \small{Dust-to-gas mass ratio} & \small{$d/g$} & \small{0.001, $\bm{0.01}$, 0.1, 1.0, 10.0, 100.0}\\
5 & \small{Central star} & \small{Sp.Type} & \small{M6, M2V, K6V, G0V, F8V, F5V, A8V, A4, A0, B9V}\\ \hline
\end{tabular}
\tablefoot{Bold numbers refer to standard model values. The central stars used in the last model series are different from the standard T~Tauri, which is a K7 spectral type.} 
\label{models}
\end{minipage}
\end{table}

\begin{figure}[htpl!]
\centering 
\vspace*{0.8cm}
\begin{minipage}[l]{0.5\textwidth}
\includegraphics[width=\textwidth]{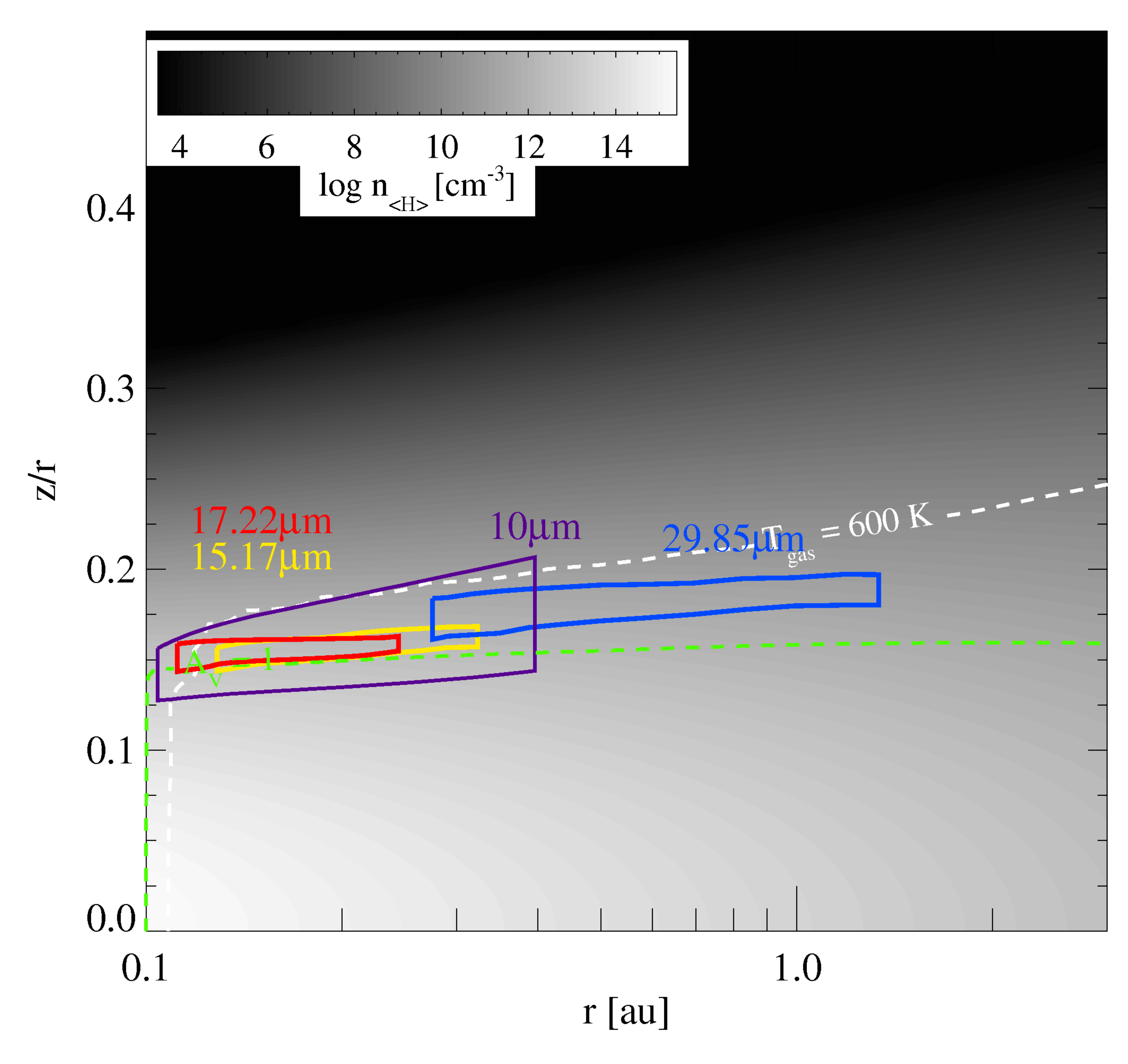}
\caption{Emitting region of the 10~$\mu$m continuum emission and the strongest water lines contributing to the 15.17~$\mu$m, 17.22~$\mu$m, 29.85~$\mu$m blend for our typical T~Tauri disk model. The emitting regions include 
15-85\% of the radial and vertical integrated flux at the given wavelengths. The green dashed line marks the optical extinction level $A_\mathrm{V}~=~1$ mag. The white dashed line marks the $T_\mathrm{gas}~=~600~$K contour.}
\label{Emitters}
\end{minipage}
\end{figure}\vspace*{0.3cm}

Fig.~\ref{Emitters} shows that these two spectral features are emitted from the same region in our standard T~Tauri model, and so they arise under the same physical and thermal conditions. Hence, we consider the 10~$\mu$m peak 
strength as an indicator of the continuum opacity in the 10-20~$\mu$m water emitting region in disks. The 29.85~$\mu$m blend is emitted from larger radii compared to the silicate feature, but there is a partial overlap, and so
the idea still holds also for this blend.

\begin{figure}[htpl!]
\begin{minipage}[l]{0.5\textwidth}
\includegraphics[width=\textwidth]{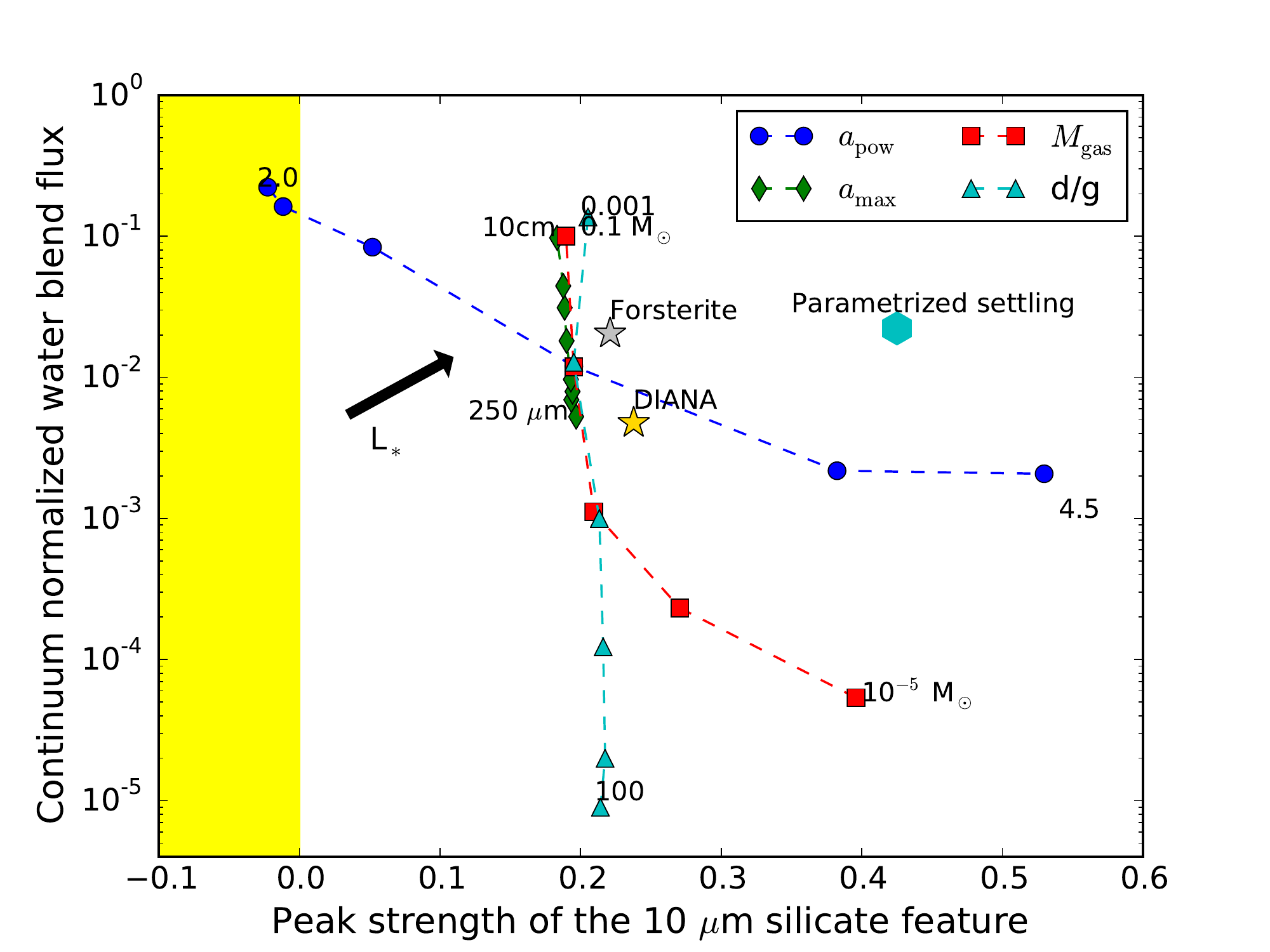}
\caption{Plot of a representative water blend to continuum ratio (15.17~$\mu$m) versus the 10~$\mu$m silicate feature for four model series. The blue dots models have different dust size power law 
distribution index. The green diamonds models have different maximum dust size. The red squares models have different gas mass. The cyan triangle models have different dust-to-gas
 ratio. 
Extreme values for each series are labeled. The shift in the plot produced by models with different $L_\mathrm{star}$ is shown by the black arrow. The yellow region marks the region of models in which the 10~$\mu$m features 
are in absorption and the peak fluxes become negative. The gold star marks the standard disk model with the DIANA opacity function, and the silver star marks the model with the fully crystalline dust opacity 
function. The cyan hexagon marks the model with the parametrized settling prescription.}
\label{Mode}
\end{minipage}
\end{figure}

Fig.~\ref{Mode} describes the results from four different model series. The first two model series we explored (1,2 in Table~\ref{models}) produce directly an effect on the abundance of $\mu$m-size grains. Extending the maximum 
dust size, maintaining constant the dust mass and the power law index of the size distribution, reduces the population of small grains. As a consequence the strength of the 10~$\mu$m feature becomes weaker. Reducing the value of
the dust size power law distribution index has the effect of putting more of the dust mass into large particles, and this again reduces the population of small grains. Finally, the effect of changing the gas mass (series 3) is 
more indirect. It is related to the radial temperature profile due to the presence/absence of large grains in the upper disk layers. If $M_\mathrm{gas}$ is low, big grains are settled to the midplane, and the small grains in the 
upper layers are more exposed to the stellar irradiation, due to the partially reduced opacity.
Our models show clearly that for a fixed dust composition, changing any of the individual disk properties (1,2,3 in Table~\ref{models}), water blend fluxes anti-correlate with the 10~$\mu$m silicate peak flux. Models with 
different luminosities of the central star (series 4), instead, show simultaneously stronger mid-IR water and silicate feature with increasing stellar luminosity (Fig.~\ref{Mode}). 

If we analyze our models with different $M_\mathrm{gas}$, we find a sharp transition at our canonical model ($M_\mathrm{gas}\!=\!0.01~$M$_{\odot}$; Fig.~\ref{Mode}). Models with larger gas mass, show stronger mid-IR water lines 
at fixed 10~$\mu$m peak strength and the blend emitting regions are moved outward by up to 1~au and slightly upward with respect to the 10~$\mu$m silicate emitting region in the disk. Models below a gas mass of 0.001~M$_{\odot}$
show no detectable mid-IR water lines (now emitted inward of 0.2~au) and a stronger 10~$\mu$m peak. 
This is due to the strong settling (because of the lack of gas support), that reduces the number of larger grains in the upper layers. In our prescription for the dust settling \citep{dubrulle}, the dust vertical scale 
height depends on the gas density. In the inner disk, grains are always unsettled in models with a significant amount of gas mass ($M_\mathrm{gas}~\ge~$0.01~M$_{\odot}$).
Only in cases of very low gas mass, the settling becomes important even in the inner disk. In this case, the emission at 10~$\mu$m becomes purely driven by small grains, which produce a much stronger emission feature. 
A similar effect can be obtained in the models by changing the amount of dust in the disk.

Our standard description for the dust opacity is based on interstellar silicates, while our target list suggests that dust composition is different in protoplanetary disks. For study the effect of a different dust 
composition, we perform two additional tests on our standard disk model using different opacity functions: fully crystalline dust grains (labelled Forsterite in Fig.~\ref{Mode}), DIANA standard opacities, \citep[60\% forsterite,
15\% amorphous carbon, 25\% vacuum; labelled DIANA in Fig.~\ref{Mode}, details can be found in][]{min2}. The two different opacity functions produce a very similar shape of the 10~$\mu$m feature, suggesting that the crystallinity 
degree alone is not able to reproduce the most extreme observations of peak strength in our sample. The increased opacity of the DIANA standard case suppresses mid-IR water line strength by about a factor two.

Given that our physical settling prescription is not able to affect the inner disk, we performed an additional test using a parametrized prescription for the settling, in which only dust grains above a certain size are settled, 
according to 

\begin{equation}\label{Sett}
c_{s}(z,a)\!=\!c_{s}(z)\cdot min(1,a/a_\mathrm{settle})^{-e_\mathrm{settle}},
\end{equation}

\noindent
where $c_{s}(z,a)$ is the sound speed, used to compute the scale height of dust of size $a$, at the height $z$ above the midplane. $a_\mathrm{settle}$ is the size threshold above which dust starts to be settled. 
$e_\mathrm{settle}$ is an exponent that modulates the size dependency of vertical settling. In our test we used $a_{settle}~=~$1$~\mu$m and a very extreme value for $e_\mathrm{settle}$, that is set to 1, four time larger than 
the value adopted in \citet{kamp3}. A different settling prescription, which affects even the inner disk, produces only only a factor two stronger peak values of the silicate feature, while mid-IR water line flux is almost
unaffected, due to the fact that smaller grains (main carrier of the continuum opacity), are still up in the disk layers from which mid-IR water lines are emitted.

\section{Mid-IR water and silicate relation}\vspace{3mm}

\begin{figure}[htpl!]
\begin{minipage}[l][12.3cm]{0.5\textwidth}
\includegraphics[width=1.1\textwidth]{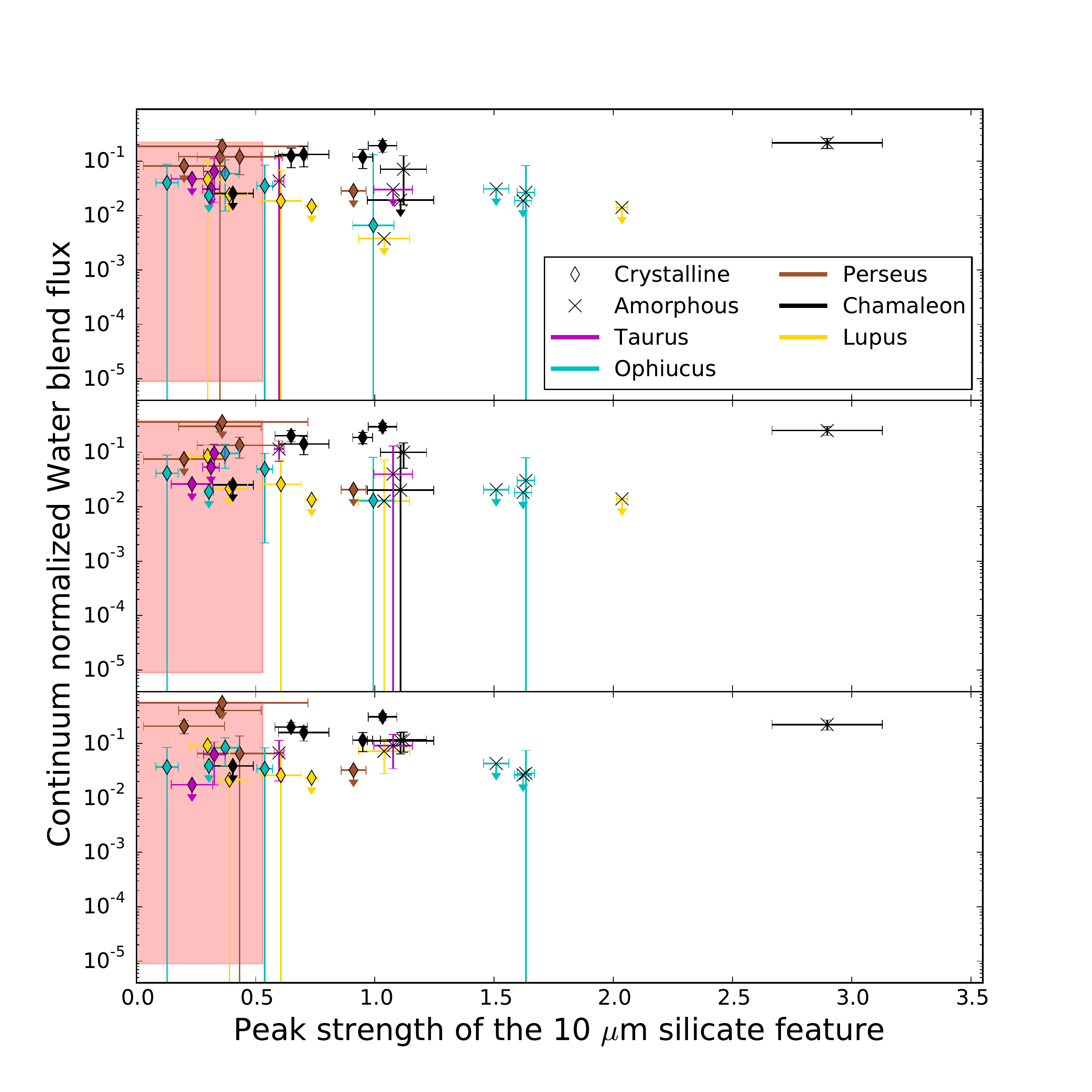}
\caption{Plots of the water blend fluxes (top 15.17~$\mu$m, middle 17.22~$\mu$m, bottom 29.85~$\mu$m) divided by the local continuum flux, versus 10~$\mu$m silicate feature peak strength. Diamonds are targets 
with crystalline dust features ($F_{11.3}/F_{9.8}\!>~$0.71), crosses are for the amorphous grains. The color coding is based on the ratio $F_{30}/F_{13}$ reported in \citet{oliveira}. 
Magenta objects are flat disks (ratio below 1.5), cyan are flared objects, yellow are cold disks and brown are edge on targets (ratio above 15). Black targets are the one for which this ratio is not available. Arrows indicates 
upper limit measurements. The shaded pink boxes encircle the area covered in all of our models.}
\end{minipage}
\label{theplot}
\end{figure}

We decided to plot our available observations for the mid-IR water line fluxes, against the computed 10~$\mu$m silicates peak strength.
Fig.~\ref{theplot} shows the water blend to continuum flux ratios, versus the 10~$\mu$m silicate peak strength, for a sample of T~Tauri star disks listed in Table~\ref{T1}, for which the Spitzer spectrum was available 
(Sect.~\ref{s1}). There is no clear correlation between mid-IR water line fluxes and the peak strength of the 10~$\mu$m silicate feature, contrary to what we find in several model series. A possible reason could be the 
diversity of the sample, since the data are from disks with different central star, dust composition and geometry.

We classify our observed sample, in order to discover possible trends with different geometry or dust properties. At first, we consider the different outer disk scale height and flaring as possible discriminator, based on the 
mid-IR slope $F_{30}/F_{13}$ and on the classification reported in \citet{oliveira}. Almost all of our targets are flared disks (cyan objects in Fig.~\ref{theplot}), with a few cases of cold disks or edge on objects.
We determine the crystallinity degree, adopting the criterium used in \citet{vanboekel,bouwman2}, that crystalline if $F_{11.3}/F_{9.8}~>~0.71$. We find that 2/3 of our targets 
have crystalline dust grains, and there is no correlation between crystallinity and disk geometry. 
Disks with crystalline silicates seem to cluster in a region of smaller silicate peak fluxes.

\noindent
The observations are unfortunately affected by large error bars for the 10~$\mu$m peak, and several upper limits for water blend fluxes (Fig.~\ref{theplot}). This is due to the sometimes low S/N of the Spitzer spectra, and it 
prevents us from clear conclusions on the presence of correlations in the data. 

\vspace*{0.5cm}

Due to the more balanced distribution of the number of targets with amorphous (10 objects) and crystalline (23 objects) dust grains, we use a basic statistical test to investigate the existence of two 
distinguished groups of objects with different mid-IR water line fluxes and silicate feature strength. 
We perform a Kolmogorov-Smirnov analysis, an Anderson-Darling and a Mantel-Haenszel statistical tests (for data with upper limits), using either R code\footnote{\tiny{$\mathrm{https://www.r-project.org/}$}} or python\footnote{\tiny{$\mathrm{http://docs.scipy.org/doc/scipy-0.16.0/reference/generated/scipy.stats}$}}. 
According to the results from these tests, we find that the 10~$\mu$m peaks do not pertain to the same distributions, but can be statistically classified in the groups of amorphous and crystalline. However, the water line
fluxes observed in disks with both type of dust grains pertain to the same distribution (Table~\ref{KS}). This result suggests that different levels of 10~$\mu$m peak strength (and so the crystallinity of dust) are not behind 
the differences in water blend strength. 
We can conclude from the Kolmogorov-Smirnov and the Mantel-Haenszel statistics that water blend fluxes at this wavelength are clearly unaffected by 10~$\mu$m peak strength.
The fact that the distributions of water emission for amorphous and  crystalline grains seem to be drawn from the same distribution indicates that the two observables are independent (water line fluxes and crystallinity). 
Since we took the 10~$\mu$m peak strength as a proxy for the local dust opacity in the water emitting region, we tentatively conclude that opacity does not directly suppress the mid-IR flux.

\begin{table}
\centering
\vspace*{1.5cm}
\begin{minipage}[l][4.2cm]{0.5\textwidth}
\caption{Kolmogorov-Smirnov, Anderson-Darling \& Mantel-Haenszel test results}
\begin{tabular}{c c c c}
\hline\hline
Data & KS & AD & MH \\ 
 & p-value$^{(*)}$ & p-value$^{(*)}$ & p-value$^{(*)}$ \\ \hline
Peak 10~$\mu$m & 6.921$\cdot$10$^{-7}$ & 2.457$\cdot$10$^{-6}$ & n.a.\\ 
Water 15.17~$\mu$m & 0.246 & 0.163 & 0.166\\
Water 17.22~$\mu$m & 0.285 & 0.189 & 0.173\\
Water 29.85~$\mu$m & 0.936 & 0.696 & 0.774\\ \hline
\end{tabular}
\tablefoot{$(*)$ p-value represents the coefficient that set if we can discard the null hypothesis (samples pertain to the same distribution if p~$>$~0.05 for KS and AD, and p~$>$~0.1 for MH.)}
\label{KS}
\end{minipage}
\end{table}

\section{Comparison between observations and models}\vspace{3mm}
\label{3}

Our model series 2 and 3 clearly predict an anticorrelation between the strength of the 10~$\mu$m silicate feature and mid-IR water line fluxes. This is however not observed in our data sample. Possible reasons for the 
apparent non correlation are the complexity of individual objects, that could be described only by a proper combination of our model series parameters. Also, our sample is very limited, and additional observations 
should be included to enlarge the sample by a factor 3 at least. This implies to get future spectra from JWST/MIRI (5-28~$\mu$m), with larger sensitivity and spatial resolution to improve the statistics and to allow the 
construction of a more homogeneous sample, e.g. focus on single spectral type, same disk mass.

Considering purely the observations, we can only confirm the presence of two distinct populations, in which disks with crystalline dust have peak strength lower than 1.1.

The range of mid-IR water line fluxes covered by the models overlaps well with the observations, but this is not the case for the 10~$\mu$m silicate feature strength. 

Our model series underestimate the value of the  10~$\mu$m peak strength, which can be up to 3 in the observations. The model series able to produce stronger peak values have very steep dust size power law distribution 
and/or very low gas content. Different opacity functions, using different degrees of crystallization, are not enough to boost the peak strength up to the extreme observations we have. As discussed in Sect.~\ref{m2}, the more 
physical 
settling prescription does not allow dust to settle in the inner disk and hence does not cover the high end of observed silicate peak strength (series 3). Adopting an alternative settling prescription, that forces dust above a 
certain size threshold to be settled, we produce a larger peak strength for our standard model.
This could indicate the presence of strong dust settling in certain disks, beyond the physical description provided by \citet{dubrulle}. 

Objects with very strong 10~$\mu$m peak show systematically lower mid-IR water line fluxes; however, there are very few objects with such high silicate peak limiting our statistics. According to our model results trend, these 
targets could have lower gas mass or a particularly high dust content in the innermost disk regions, where planet formation occurs. Disks in which planet formation happens, can also experience amorphization of dust grains 
\citep{watson2}, this is consistent with the fact that in our data sample, disks with amorphous dust grains have stronger 10~$\mu$m peak strength.

To date, ALMA high spatial resolution observations could investigate, if some of the disks with strong water emission are indeed dust depleted in the inner 10~au.

\section{Conclusions}\vspace{3mm}
\label{5}

Some of our model series (1,2,3,4) suggest the presence of a regular trend between the strength of the 10~$\mu$m silicate dust feature and the mid-IR water blend fluxes. This is directly related to changing dust opacity. 
A stronger silicate feature is produced by disk properties such as very small dust grains, very steep power law dust size distribution, very low gas mass. It is an indicator of larger continuum opacity and leads to weaker mid-IR 
water lines. 
Both dust grains and water behave in the same manner with respect to the stellar bolometric luminosity (series 5), since both gas and (in particular) dust get warmer.  

Observational samples are likely more heterogeneous than what our models currently describe. Disks around T~Tauri stars show mainly processed and more crystalline dust, while our modeling is performed with ISM dust (mainly 
amorphous)
. The two alternative cases of dust opacity function we explored, cannot explain observed peak strength larger than 1. A test performed with a parametric description of settling, shows that our models would be systematically
shifted to a peak strength larger of about 0.2, if the inner disk would be settled.

Observations are consistent with relatively flared disks around T~Tauri stars and crystalline dust features. 
From statistical tests, we found that the peak strength is different in disks with amorphous and crystalline grains. We also found that the water line fluxes in the two samples of amorphous and crystalline silicate disks are 
indistinguishable. This suggests that dust opacity due to the different composition does not drive the difference in mid-IR water fluxes. The independent behavior of these two observables, and the results from models of 
extremely settled disks, leave an alternative explanation, namely that some disks have an inner ring within 10~au enhanced in dust or depleted in gas, experiencing strong settling. High spatial resolution 
observations (ALMA, JWST) are needed to confirm this hypothesis.

\begin{acknowledgements}
We thank Aaron Greenwood for the language editing and precious suggestions about how to improve the clarity of the paper. We thank Klaus Pontoppidan and Jeroen Bouwman for productive and fruitful discussions on the topic of 
this paper. The research leading to these results has received funding from the European Union Seventh Framework Programme FP7-2011 under grant agreement no 284405.
\end{acknowledgements}

\bibliographystyle{aa}
\bibliography{Silicate}


\appendix

\section{Observational data}

Observations used in this study (Table~\ref{T1}).

\begin{table*}
\centering
\caption{Selected T~Tauri disk system}
\begin{tabular}{p{2.cm}p{2.cm}p{2.2cm}p{2.2cm}p{2.2cm}cccc}
\hline\hline
Source name & Peak 10~$\mu$m & 15.17~$\mu$m$^{(1)}$  & 17.22~$\mu$m$^{(1)}$ & 29.85~$\mu$m$^{(1)}$  & 11/9 & 30/13$^{(2)}$ & $L_\mathrm{star}$ & Si feature \\ 
 & & \small{[10$^{-14}$~erg~cm$^{-2}$s$^{-1}$]} & \small{[10$^{-14}$~erg~cm$^{-2}$s$^{-1}$]} & \small{[10$^{-14}$~erg~cm$^{-2}$s$^{-1}$]} & & & \small{[$L_{\odot}$]} & (Cry/Amo)$^{(3)}$\\ \hline
LkHa270 &  0.20$\pm$0.17 & $<$1.15 & $<$1.08 & 2.62$\pm$0.22 & 1.55 & 1.59 & ... & Cry\\
 LkHa326 & 0.43$\pm$0.18 & 1.50$\pm$0.16 & 1.98$\pm$0.16 & 0.96$\pm$0.13 & 1.15 & 2.88 & 0.05$^{(4)}$ & Cry\\
 LkHa327 & 0.35$\pm$0.17 & 3.19$\pm$0.86 & 7.75$\pm$0.80 & 7.37$\pm$0.48 & 1.66 & 1.23 & ... & Cry\\
 AS205 &  0.54$\pm$0.03 & 9.21$\pm$0.52 & 14.50$\pm$0.59 & 7.46$\pm$0.38 & 0.72 &  1.76 & 7.10$^{(5)}$ & Cry\\
 EXLup & 1.04$\pm$0.11 & $<$0.42 & 1.47$\pm$0.14 & 4.65$\pm$0.08 & 0.52 & 1.56 & 0.39$^{(5)}$ & Amo\\
 GQLup & 0.30$\pm$0.07 & 0.82$\pm$0.07 & 1.60$\pm$0.07 & 1.53$\pm$0.05 & 0.86 & 2.04 & 0.80$^{(5)}$ & Cry\\
 HTLup & 0.39$\pm$0.08 & $<$2.01 & $<$1.98 & 1.34$\pm$0.17 & 0.85 & 1.53 & 1.45$^{(5)}$ & Cry\\
 RNO90 & 0.37$\pm$0.06 & 4.65$\pm$0.19 & 8.05$\pm$0.20 & 4.67$\pm$0.11 & 0.71 & 1.73 & 4.06$^{(6)}$ & Cry\\
 RULup & 0.61$\pm$0.08 & 2.87$\pm$0.15 & 4.01$\pm$0.16 & 2.77$\pm$0.09 & 0.93 & 1.94 & 0.42$^{(5)}$ & Cry\\
 V1121Oph & 0.99$\pm$0.09 & 0.89$\pm$0.24 & 2.14$\pm$0.26 & 2.02$\pm$0.14 & 0.72 & 1.87 & 1.50$^{(7)}$ & Cry\\
 VWCha & 0.95$\pm$0.04 & 3.29$\pm$0.12 & 5.85$\pm$0.13 & 2.99$\pm$0.07 & 0.87 & 2.08 & 2.34$^{(5)}$ & Cry\\
 VZCha & 0.65$\pm$0.07 & 1.55$\pm$0.08 & 2.30$\pm$0.08 & 1.24$\pm$0.03 & 0.72 & 1.03 & 0.46$^{(5)}$ & Cry\\
 WaOph6 & 0.13$\pm$0.05 & 1.25$\pm$0.06 & 1.23$\pm$0.06 & 0.84$\pm$0.04 & 1.29 & 1.66 & 0.67$^{(5)}$ & Cry\\
 DRTau & 0.33$\pm$0.05 & 4.53$\pm$0.19 & 7.13$\pm$0.19 & 3.73$\pm$0.10 & 0.72 & ... & 2.50$^{(5)}$ & Cry\\
 SXCha & 1.12$\pm$0.10 & 1.29$\pm$0.10 & 2.00$\pm$0.10 & 1.67$\pm$0.07 & 0.63 & 1.72 & 0.44$^{(5)}$ & Amo\\
 SYCha & 1.11$\pm$0.14 & $<$0.15 & 0.20$\pm$0.05 & 0.71$\pm$0.03 & 0.67 & 2.16 & 0.37$^{(5)}$ & Amo\\
 TWCha & 2.90$\pm$0.23 & 1.01$\pm$0.03 & 1.72$\pm$0.05 & 0.91$\pm$0.03 & 0.59 & 2.07 & 0.90$^{(5)}$ & Amo\\
 WXCha & 1.03$\pm$0.06 & 1.80$\pm$0.07 & 2.66$\pm$0.07 & 1.49$\pm$0.03 & 0.83 & 0.99 & 0.68$^{(5)}$ & Cry\\
 XXCha & 0.70$\pm$0.11 & 0.66$\pm$0.05 & 0.76$\pm$0.05 & 0.66$\pm$0.03 & 0.79 & 1.70 & 0.26$^{(5)}$ & Cry\\
 LkCa8 & 1.08$\pm$0.08 & $<$0.17 & 0.27$\pm$0.05 & 0.71$\pm$0.06 & 0.59 & 2.28 & ... & Amo\\
 Haro1-16 & 1.63$\pm$0.04 & 0.66$\pm$0.06 & 1.24$\pm$0.06 & 1.15$\pm$0.05 & 0.59 & 4.30 & 2.00$^{(5)}$ & Amo\\
 AATau & 0.60$\pm$0.02 & 0.48$\pm$0.06 & 1.48$\pm$0.06 & 0.74$\pm$0.03 & 0.69 & ... & 0.98$^{(5)}$ & Amo\\
 LkHa271 & 0.36$\pm$0.36 & $<$0.45 & $<$1.05 & $<$2.1 & 1.21 & 2.54 & ... & Cry\\
 LkHa330 & 0.91$\pm$0.05 & $<$0.73 & $<$0.92 & $<$3.12 & 1.02 & 11.74 & ... & Cry\\
 V710Tau & 0.31$\pm$0.03 & $<$0.17 & $<$0.32 & $<$0.69 & 0.81 & 1.41 & ... & Cry\\
 CokuTau4 & 0.23$\pm$0.09 & $<$0.29 & $<$0.33 & $<$0.39 & 0.95 & 16.96 & ... & Cry\\
 Sz50 & 0.40$\pm$0.09 & $<$0.20 & $<$0.22 & $<$0.30 & 1.75 & 2.31 & ... & Cry\\
 GULup & 0.45$\pm$0.06 & $<$0.20 & $<$0.21 & $<$0.39 & 0.76 & ... & ... & Cry\\
 IMLup & 0.74$\pm$0.03 & $<$0.28 & $<$0.28 & $<$0.34 & 0.76 & 1.85 & ... & Cry\\
 RYLup & 2.04$\pm$0.02 & $<$0.75 & $<$0.85 & $<$1.11 & 0.50 & 3.92 & ... & Amo\\
 Haro1-1 & 1.51$\pm$0.05 & $<$0.18 & $<$0.18 & $<$0.53 & 0.59 & 4.70 & ... & Amo\\
 Haro1-4 & 1.62$\pm$0.04 & $<$0.30 & $<$0.38 & $<$0.60 & 0.59 & 2.44 & ... & Amo\\
 Haro1-17 & 0.30$\pm$0.02 & $<$0.11 & $<$0.11 & $<$0.28 & 0.90 & 1.96 & ... & Cry\\
\hline
\end{tabular}
\tablefoot{(1) Fluxes from \citet{pontoppidan2}, scaled at 140 pc of distance. (2) Values taken from \citet{olofsson}. (3) Cry: crystalline silicate feature; Amo: amorphous silicate feature. (4) Luminosities from \citet{rigliaco}. (5) Luminosities from \citet{salyk2}.
(6) Luminosities from \citet{salyk4}. (7) Luminosities from \citet{blevins}.}
\label{T1}
\end{table*}

\end{document}